\documentclass[conference, onecolumn]{IEEEtran}
\usepackage[T1]{fontenc}
\usepackage[latin9]{inputenc}
\usepackage{calc}
\usepackage{amsmath}
\usepackage[unicode=true,
 bookmarks=true,bookmarksnumbered=true,bookmarksopen=true,bookmarksopenlevel=1,
 breaklinks=false,pdfborder={0 0 0},pdfborderstyle={},backref=false,colorlinks=false]
 {hyperref}
\hypersetup{pdftitle={Your Title},
 pdfauthor={Your Name},
 pdfpagelayout=OneColumn, pdfnewwindow=true, pdfstartview=XYZ, plainpages=false}

\makeatletter
\usepackage[caption=false,font=footnotesize]{subfig}
\usepackage[american]{babel}
\usepackage{cuted}
\usepackage{color}

\makeatother

\begin{document}
\title{Clock Offsets and Ranges \\
from Optical Pseudoranges}
\author{
\IEEEauthorblockN{Christoph Günther}
\IEEEauthorblockA{
	Technical University of Munich (TUM)\\ 
	Theresienstraße 90\\
	D-80333 München\\
	Email: christoph@guenther.space\\
	October 30, 2026
}
}
\maketitle
\begin{abstract}
This note describes the estimation of ranges and clock offsets from
measured pseudoranges in the context of the Kepler Navigation System.
The latter system continuously adapts the frequency of its clocks
to ensure that they all show the same time independently of where
they are. In Medium Earth Orbits, the joint problem of estimating
ranges and clock offsets can then be  factorized.
\end{abstract}

\IEEEpeerreviewmaketitle{}

\section{Introduction\label{sec:Introduction}}

Future satellite navigation systems will use optical intersatellite
links to synchronize their clocks, measure inter-satellite ranges
and to exchange information. Synchronization and ranging may be based
on pseudoranges such as those used in Global Navigation Satellite
Systems (GNSS). The main difference is, however, that the chip rates
are much higher, e.g. 25 Gcps for a prototype built at DLR (Surof
et al. \cite{SurPolWolMat26}). The acquisition of the optical signal
has become state of the art and is implemented in terminals of TESAT,
Mynaric and others. After acquisition, the optical carrier phase is
tracked to enable the demodulation of the signals. Different options
exist and have been implemented. In the Kepler System proposed by
Günther \cite{Guen18}, and described in Giorgi et al. \cite{GioSchTraMat19},
the time references on the satellites are Ultra Stable Oscillators
(USO). Such devices exists in space qualified versions. The tie to
Time Atomic International (TAI) and thus Universal Time Coordinated
(UTC) is established by an Active Hydrogen Maser (AHM), which provides
a real-time representation of TAI. The synchronization (and ranging)
in the system is established by evaluating the timing of the modulation
only. 

The Kepler System enables highly accurate orbits based on inter-satellite
ranges (Michalak et al. \cite{MicDasTesGio22}) as well as an extremely
precise synchronization (Trainotti et al. \cite{TraDasGioKho22},
\cite{TraDasGioKho23} and Günther \cite{Guen26}). The latter synchronization
can furthermore be made accurate by calibration. The system becomes
then free of biases (Günther \cite{GueMic26}). The performance in
ranging and in synchronization relies on the availability of precise
ranges and/or clock offsets. They need to be somehow extracted from
measured pseudoranges. Relativity complicates the task, since any
free-running clocks would be perceived as running at a different pace
when observed from a different orbit. Fortunately, the Kepler System
has all its clocks, including the USOs on the satellites and the AHM
on the ground, always showing the same time, thanks to a continuous
adaptation of their frequency. The derivation in this note is based
on this assumption. 

\section{Clock Offsets and Ranges from Pseudoranges\label{sec:Ranges-and-Clock}}

In the present paper, we study the clock offset 
\begin{equation}
\delta^{kl}(t)=\delta^{k}(t)-\delta^{l}(t)\label{eq:Delta_delta}
\end{equation}
between the clocks on Satellite $k$ and $l$ and the range between
them at time $t$:
\begin{equation}
r^{kl}(t)=\left\Vert \vec{r}^{\,k}(t)-\vec{r}^{\,l}\big(t\big)\right\Vert .\label{eq:r^kl_t}
\end{equation}
The vectors $\vec{r}^{\,k}(t)$ and $\vec{r}^{\,l}(t)$ in Equation
(\ref{eq:r^kl_t}) are the positions of Satellite $k$ and $l$, respectively.
Equation (\ref{eq:Delta_delta}), suggests the existence of some reference
time, which we shall call Kepler System Time (KST), see Trainotti
et al. \cite{TraDasGioKho22}, Trainotti \cite{Trai26} and Günther
\cite{Guen26}. Since the clock offsets vary very slowly, their dependence
on $t$ is typically not made explicit.

Optical pseudoranges $\rho^{kl}$ are defined as the time needed for
the optical signal to propagate from on satellite to the other, with
time measured using the clocks of the receiving and transmitting satellite
respectively:

\begin{align}
\rho^{kl} & =c(t_{R}^{k}-t_{T}^{l})\label{eq:rho^kl_def}\\
 & =\left\Vert \vec{r}^{\,k}(t_{R}^{k})-\vec{r}^{\,l}\big(t_{T}^{l}\big)\right\Vert +c\delta^{kl}.\nonumber 
\end{align}
In this expression, $t_{R}^{k}$ is a time tag associated with a particular
phasing of the signal received by Satellite $k$ measured with the
clock of Satellite $k$, which was transmitted by Satellite $l$ when
its clock was showing $t_{T}^{l}.$ The beginning of a spreading code
in a particular frame is a typical example for such a time tag. The
second line of Equation (\ref{eq:rho^kl_def}) is obtained by mapping
the clocks of the satellites to the common reference KST and by substituting
the range once the times have been mapped. The latter assumes a free-space
propagation model, which is valid, since the signals propagate in
vacuum and since the Shapiro delay is negligible \cite{DasGio21}.
The aim of the next step is to relate the pseudorange $\rho^{kl}$
to the range $r^{kl}$ at the given time $t_{R}^{k}$, i.e.: 

\begin{equation}
r^{kl}=\left\Vert \vec{r}^{\,k}(t_{R}^{k})-\vec{r}^{\,l}\big(t_{R}^{k}\big)\right\Vert ,\label{eq:r^kl}
\end{equation}
with $r^{kl}=r^{kl}(t_{R}^{k})$. Creating such a relationship requires
knowledge of the average $\vec{v}_{f}$ of the velocity $\vec{v}^{\,l}(s)$
of Satellite $l$ during the forward transmission $s\in[t_{T}^{l}-\delta^{l},t_{R}^{k}-\delta^{k}]$,
with the integration variable $s$ being referred to KST: 
\begin{equation}
\vec{v}_{f}=\frac{1}{t_{R}^{k}-t_{T}^{l}-\text{\ensuremath{\delta^{kl}}}}\int_{t_{T}^{l}-\delta^{l}}^{t_{R}^{k}-\delta^{k}}ds\,\vec{v}^{\,l}(s).\label{eq:vec=00007Bv=00007D_f}
\end{equation}
Equation (\ref{eq:rho^kl_def}) may be rewritten in the form:
\begin{equation}
t_{T}^{l}=t_{R}^{k}-\frac{\rho^{kl}}{c},\label{eq:t_T^l}
\end{equation}
which implies a non-linear relationship. Equation (\ref{eq:vec=00007Bv=00007D_f})
implicitly assumes that the offset $\delta^{kl}$ is known, while
it is still to be determined. The average velocity $\vec{v}_{f}$
can then be computed whenever $\vec{v}^{\,l}(s)$ is known. The velocity
$\vec{v}^{\,l}(s)$ is itself determined by the orbital movement of
Satellite $l$, i.e. primarily by earth gravity and to a smaller extent
by tidal forces from other celestial bodies and by the solar pressure.
There is no way for Satellite $l$ to estimate this information from
isolated relative measurements with respect to other satellites. It
must thus be retrieved from the orbit estimation. The excellent predictability
of the orbits of the proposed Kepler navigation system makes this
information available, at least after an initialization phase. Dassié
et al. showed that errors in the velocity estimate in a comparable
context are below 0,5 mm/s. The mutual dependency of the variables
in the above split approach is a limiting factor in orbit determination
and time synchronization. It must be assessed more deeply by simulations
in the future. A potential limitation would need to be overcome in
a joint estimation of all relevant variables, which is rather complex,
however.

The Equations (\ref{eq:rho^kl_def}), (\ref{eq:r^kl}) and (\ref{eq:vec=00007Bv=00007D_f})
are the basis for establishing a relation between the pseudorange
$\rho^{kl}$, the range $r^{kl}$ and the clock offset $\delta^{kl}:$
\begin{align}
\rho^{kl} & =\left\Vert \vec{r}^{\,k}(t_{R}^{k})-\vec{r}^{\,l}(t_{R}^{k})+\vec{v}_{f}\big(t_{R}^{k}-t_{T}^{l}-\text{\ensuremath{\delta^{kl}}}\big)\right\Vert +c\delta^{kl}\label{eq:rho^kl}\\
 & =\left((r^{kl})^{2}+2r^{kl}\left(\vec{e}^{\,kl}\right)^{T}\vec{v}_{f}(t_{R}^{k}-t_{T}^{l}-\text{\ensuremath{\delta^{kl}}})+v_{f}^{2}(t_{R}^{k}-t_{T}^{l}-\text{\ensuremath{\delta^{kl}}})^{2}\right)^{1/2}+c\delta^{kl}\nonumber 
\end{align}
with $v_{f}=\|\vec{v}_{f}\|$, and with
\[
\vec{e}^{\,kl}=\frac{\vec{r}^{\,k}(t_{R}^{k})-\vec{r}^{\,l}(t_{R}^{k})}{\left\Vert \vec{r}^{\,k}(t_{R}^{k})-\vec{r}^{\,l}(t_{R}^{k})\right\Vert }
\]
being the directional vector pointing from $\vec{r}^{\,l}(t_{R}^{k})$
to $\vec{r}^{\,k}(t_{R}^{k})$. Let furthermore:
\[
\tilde{v}_{f}=\left(\vec{e}^{\,kl}\right)^{T}\vec{v}_{f},
\]
be the projection of the velocity vector $\vec{v}_{f}$ onto the directional
vector $\vec{e}^{\,kl}$ and let 
\[
z_{f}=c\frac{t_{R}^{k}-t_{T}^{l}-\delta^{kl}}{r^{kl}}=\frac{\rho^{kl}-c\delta^{kl}}{r^{kl}}
\]
then Equation (\ref{eq:rho^kl}) may be rewritten in the form
\begin{equation}
1+2\frac{\tilde{v}_{f}}{c}z_{f}+\left(\left(\frac{v_{f}}{c}\right)^{2}-1\right)z_{f}^{2}=0,\label{eq:z_f-equ}
\end{equation}
which has the relevant solution:
\begin{align}
z_{f} & =\sqrt{1+\left(\frac{\tilde{v}_{f}}{c}\right)^{2}-\left(\frac{v_{f}}{c}\right)^{2}}-\frac{\tilde{v}_{f}}{c}\label{eq:z_f-sol}\\
 & =1-\frac{\tilde{v}_{f}}{c}+\frac{1}{2}\left(\frac{\tilde{v}_{f}}{c}\right)^{2}-\frac{1}{2}\left(\frac{v_{f}}{c}\right)^{2}+O\left(\frac{v_{f}}{c}\right)^{4}.\nonumber 
\end{align}
The other solution would be associated with the unphysical situation
in which $\rho^{kl}$ and $r^{kl}$ have opposite signs! Equation
(\ref{eq:z_f-sol}) provides a relation between the measured pseudorange
$\rho^{kl}$ and the unknown range $r^{kl}$ and clock offset $\delta^{kl}.$
Using $\delta^{lk}=\delta^{l}-\delta^{k}=-\delta^{kl}$, a second
relation is obtained by considering the return link under the simplifying
but unessential assumption of not delay between reception and transmission
$t_{T}^{k}=t_{R}^{k}$:
\begin{align}
\rho^{lk} & =c(t_{R}^{l}-t_{T}^{k})\label{eq:rho^lk}\\
 & =\left\Vert \vec{r}^{\,l}(t_{R}^{l})-\vec{r}^{\,k}\big(t_{T}^{k}\big)\right\Vert +c\delta^{lk}\nonumber \\
 & =\left\Vert \vec{r}^{\,k}(t_{R}^{k})-\vec{r}^{\,l}(t_{R}^{k})-\vec{v}_{r}\big(t_{R}^{l}-t_{T}^{k}+\delta^{kl}\big)\right\Vert -c\delta^{kl}\nonumber \\
 & =\left((r^{kl})^{2}-2r^{kl}\left(\vec{e}^{\ kl}\right)^{T}\vec{v}_{r}(t_{R}^{l}-t_{T}^{k}+\delta^{kl})+v_{r}^{2}(t_{R}^{l}-t_{T}^{k}+\delta^{kl})^{2}\right)^{1/2}-c\delta^{kl},\nonumber 
\end{align}
with
\[
\vec{v}_{r}=\frac{1}{t_{R}^{l}-t_{T}^{k}+\delta^{kl}}\int_{t_{T}^{k}-\delta^{k}}^{t_{R}^{l}-\delta^{l}}ds\,\vec{v}^{\,l}(s),\quad\text{and\ensuremath{\quad}}v_{r}=\|\vec{v}_{r}\|.
\]
This equation was designed to express its result using the same range
$r^{kl}$ measured at time $t_{R}^{k}=t_{T}^{k}.$ The expression
again uses a displacement of the same Satellite $l,$ which makes
the analysis easier. The average velocities $\vec{v}_{r}$ and $\vec{v}_{f}$
only differ by the integration intervals. The interval associated
with the forward link lies before $t_{R}^{k}-\delta^{k}$ and the
one associated with the return link after $t_{T}^{k}-\delta^{k}=t_{R}^{k}-\delta^{k}.$
The definitions of $r^{kl}$, $\delta^{kl}$ and $\vec{e}^{\,kl}$
remain unchanged in the return direction as compared to the forward
direction, i.e. Equation (\ref{eq:rho^kl}). Defining 
\[
\tilde{v}_{r}=\left(\vec{e}^{\,kl}\right)^{T}\vec{v}_{r},\quad\text{and\ensuremath{\quad z_{r}=c\frac{t_{R}^{l}-t_{T}^{k}+\delta^{kl}}{r^{kl}}=\frac{\rho^{lk}+c\delta^{kl}}{r^{kl}}},}
\]
leads to the following equivalent of Equation (\ref{eq:z_f-equ}):
\[
1-2\frac{\tilde{v}_{r}}{c}z_{r}+\left(\left(\frac{v_{r}}{c}\right)^{2}-1\right)z_{r}^{2}=0,
\]
with the relevant solution
\begin{align*}
z_{r} & =\sqrt{1+\left(\frac{\tilde{v}_{r}}{c}\right)^{2}-\left(\frac{v_{r}}{c}\right)^{2}}+\frac{\tilde{v}_{r}}{c}.\\
 & =1+\frac{\tilde{v}_{r}}{c}+\frac{1}{2}\left(\frac{\tilde{v}_{r}}{c}\right)^{2}-\frac{1}{2}\left(\frac{v_{r}}{c}\right)^{2}+O\left(\frac{v_{r}}{c}\right)^{4}.
\end{align*}
The values of $z_{f}$ and $z_{r}$ are thus completely defined by
the average velocities of Satellite $l$ in the intervals $[t_{T}^{l}-\delta^{l},t_{R}^{k}-\delta^{k}]$
and $[t_{T}^{k}-\delta^{k},t_{R}^{l}-\delta^{l}]$. Using the definition
of $z_{f}$ and of $z_{r}$ in terms of $\rho^{kl},$ $\rho^{lk},$
$r^{kl},$ and $\delta^{kl},$ implies:

\begin{align*}
\rho^{kl}-c\delta^{kl} & =z_{f}r^{kl}\\
\rho^{lk}+c\delta^{kl} & =z_{r}r^{kl}
\end{align*}
and thus

\begin{align}
r^{kl} & =\frac{\rho^{kl}+\rho^{lk}}{z_{f}+z_{r}}\label{eq:r^kl_delta^kl}\\
c\delta^{kl} & =\frac{z_{r}\rho^{kl}-z_{f}\rho^{lk}}{z_{f}+z_{r}}.\nonumber 
\end{align}
As a consequence, the desired range $r^{kl}$ and clock offset $\delta^{kl}$
can be evaluated as linear combinations of $\rho^{kl}$ and $\rho^{lk}$
with coefficients that depend on $v_{f},$ $\tilde{v}_{f},$ $v_{r},$
and $\tilde{v}_{r}.$ The Equations (\ref{eq:r^kl_delta^kl}) describe
a purely geometric relation. The measurements of $\rho^{kl}$ and
$\rho^{lk}$ are affected by noise. The level of that noise is at
picosecond level, as could be established in the experiments performed
by Surof et al. \cite{SurPolMatRic19,SurPolWolMac22}. Since the noise
in the pseudorange measurements in the forward and return link are
independent and since the approximation $z_{f}\sim z_{r}\sim1$ holds
for noise considerations, the level of noise in the estimation of
the range and clock offset is very much the same as in the pseudorange
measurements.

Next consider the impact of the velocity on $z_{f}$ and $z_{r}$
more accurately. In this context, Satellite $l$ can be considered
to be on a circular orbit $\vec{r}(t)=a\left(\cos(\omega t),\sin(\omega t\right)$
with a radius $a$, when represented in its orbital plane. This is
a good approximation for Galileo, since $e=0.001$! The velocity in
that orbit is $\vec{v}(t)=\dot{\vec{r}}(t)=a\omega\left(-\sin(\omega t),\cos(\omega t)\right)$.
The period of the orbit is given by Kepler's third law $\omega=2\pi/T=\sqrt{GM/a^{3}}.$
Thus, $v(t)/c=\sqrt{GM/ac^{2}}=1.2\cdot10^{-8}.$ As a consequence,
terms with factors of $(v/c)^{2}=1.5\cdot10^{-16}$ can be neglected
in $z_{f}$ and $z_{r}$. Therefore, 
\[
z_{f}\simeq1-\frac{\tilde{v}_{f}}{c}\quad\text{and \ensuremath{\quad z_{r}\simeq1+\frac{\tilde{v}_{r}}{c}.}}
\]
The impact of the satellite velocity $\vec{v}^{\,l}$ on $z_{f}$
and $z_{r}$ is thus of the order $v/c$. The range depends on 
\[
\begin{aligned}z_{f}+z_{r}-2 & =\frac{1}{c}(\vec{v}_{r}-\vec{v}_{f})\simeq\frac{1}{c}\left(\vec{v_{l}}(t+\Delta t/2)-\vec{v}_{l}(t-\Delta t/2)\right)\\
 & \simeq\frac{\Delta t}{c}\dot{\vec{v}}_{l}(t)=\frac{-\omega^{2}\Delta t}{c\left(1-e\cos E(t)\right)^{3}}\left(\begin{array}{c}
a\cos E(t)-ae\\
b\sin E(t)
\end{array}\right)
\end{aligned}
\]
with the last expression being expressed by coordinates in the orbital
plane of the satellite. Since $\omega^{2}a/c=GM/(a^{2}c)=1.5\cdot10^{-12}\text{ [s]}^{-1}$,
and since $\Delta t\simeq75$ ms the overall expression is $1.14\cdot10^{-13},$
which means that the range $r^{kl}$ is basically the average of the
pseudoranges. This is fortunate since ranges are essential for the
estimation of orbits and thus velocities. As soon as this estimation
process has converged, accurate estimates of $\vec{v}^{\,l}$ become
the basis for accurate estimates of the time offsets $\delta^{kl}$.
Equation (\ref{eq:r^kl_delta^kl}) thus leads to a certain decoupling
of orbit and clock offset estimation, which is much simpler than a
full joint estimation.

\section{Conclusion\label{sec:Outlook-and-Conclusions}}

This note creates a link between measured pseudoranges and ranges
as well as clock offsets as needed in the algorithms of satellite
navigation. It shows that in the MEO orbits the separation can be
achieved in an iterative and interleaved manner without having to
solve the more complex joint problem.

\bibliographystyle{IEEEtran}
\bibliography{IEEEabrv,IEEEexample,../Bibliography}

\begin{thebibliography}{10}
\providecommand{\url}[1]{#1}
\csname url@samestyle\endcsname
\providecommand{\newblock}{\relax}
\providecommand{\bibinfo}[2]{#2}
\providecommand{\BIBentrySTDinterwordspacing}{\spaceskip=0pt\relax}
\providecommand{\BIBentryALTinterwordstretchfactor}{4}
\providecommand{\BIBentryALTinterwordspacing}{\spaceskip=\fontdimen2\font plus
\BIBentryALTinterwordstretchfactor\fontdimen3\font minus
  \fontdimen4\font\relax}
\providecommand{\BIBforeignlanguage}[2]{{%
\expandafter\ifx\csname l@#1\endcsname\relax
\typeout{** WARNING: IEEEtran.bst: No hyphenation pattern has been}%
\typeout{** loaded for the language `#1'. Using the pattern for}%
\typeout{** the default language instead.}%
\else
\language=\csname l@#1\endcsname
\fi
#2}}
\providecommand{\BIBdecl}{\relax}
\BIBdecl

\bibitem{SurPolWolMat26}
J.~Surof, J.~Poliak, R.~Wolf, R.~Mata~Calvo, L.~Bl{\"u}mel, and C.~G{\"u}nther,
  ``{Precise Time Transfer and Ranging for Next-Generation GNSS},'' \emph{GPS
  Solutions}, vol.~30, no.~2, p. 101, 2026.

\bibitem{Guen18}
C.~G{\"u}nther, ``{Kepler - Satellite Navigation without Clocks and Ground
  Infrastructure},'' in \emph{Proc. ION GNSS+ 2018}.\hskip 1em plus 0.5em minus
  0.4em\relax ION, 2018.

\bibitem{GioSchTraMat19}
G.~Giorgi, T.~D. Schmidt, C.~Trainotti, R.~Mata-Calvo, C.~Fuchs, M.~M. Hoque,
  J.~Berdermann, J.~Furthner, C.~G{\"u}nther, T.~Schuldt \emph{et~al.},
  ``{Advanced Technologies for Satellite Navigation and Geodesy},''
  \emph{Advances in Space Research}, vol.~64, no.~6, pp. 1256--1273, 2019.

\bibitem{MicDasTesGio22}
G.~Michalak, M.~Dassi{\'e}, A.~Testa, G.~Giorgi, C.~G{\"u}nther, and K.~H.
  Neumayer, ``{Signal-in-Space Range Error Analysis of the Simulated Broadcast
  Ephemerides for the Kepler System.}'' pp. 2838--2847, 2022.

\bibitem{TraDasGioKho22}
C.~Trainotti, M.~Dassi{\'e}, G.~Giorgi, A.~Khodabandeh, and C.~G{\"u}nther,
  ``{Autonomous Satellite System Synchronization Schemes via Optical Two-Way
  Time Transfer and Distributed Composite Clock},'' in \emph{Proceedings of the
  35th International Technical Meeting of the Satellite Division of The
  Institute of Navigation (ION GNSS+ 2022)}, 2022, pp. 3646--3661.

\bibitem{TraDasGioKho23}
------, ``{Autonomous Synchronization of Satellite Constellations via Optical
  Inter-Satellite Links},'' in \emph{2023 Joint Conference of the European
  Frequency and Time Forum and IEEE International Frequency Control Symposium
  (EFTF/IFCS)}.\hskip 1em plus 0.5em minus 0.4em\relax IEEE, 2023, pp. 1--6.

\bibitem{Guen26}
\BIBentryALTinterwordspacing
C.~G{\"u}nther, ``{Kepler System Time},'' in \emph{{European Navigation
  Conference ENC 2026} = slides, full paper in preparation}. [Online].
  Available: \url{https://cloud.1ho.st/s/WpiDfepqHpcfyEb}
\BIBentrySTDinterwordspacing

\bibitem{GueMic26}
C.~G{\"u}nther and G.~Michalak, ``{Optical Time Biases in the Kepler System},''
  in \emph{{IEEE Navicon 2026}}, 2026.

\bibitem{Trai26}
C.~Trainotti, ``{Synchronization of Next Generation GNSS},'' Ph.D.
  dissertation, Technical University of Munich, Munich, Germany, 2026.

\bibitem{DasGio21}
M.~Dassi{\'e} and G.~Giorgi, ``{Relativistic Modeling for Accurate Time
  Transfer via Optical Inter-Satellite Links},'' \emph{Aerotecnica Missili \&
  Spazio}, vol. 100, no.~3, pp. 277--288, 2021.

\bibitem{SurPolMatRic19}
J.~Surof, J.~Poliak, R.~M. Calvo, M.~Richerzhagen, R.~Wolf, and T.~Schmidt,
  ``{Laboratory Characterization of Optical Inter-Satellite Links for Future
  GNSS},'' in \emph{Proceedings of the 32nd International Technical Meeting of
  The Satellite Division of the Institute of Navigation (ION GNSS+ 2019)},
  2019, pp. 1051--1058.

\bibitem{SurPolWolMac22}
J.~Surof, J.~Poliak, R.~Wolf, L.~Macri, R.~M. Calvo, L.~Bl{\"u}mel,
  P.~Dominguez, G.~Giorgi, L.~Agazzi, J.~Furthner \emph{et~al.}, ``{Validation
  of Kepler Time and Frequency Transfer on a Terrestrial range of 10.45 km},''
  in \emph{Proceedings of the 35th International Technical Meeting of the
  Satellite Division of The Institute of Navigation ({ION GNSS+ 2022})}, 2022,
  pp. 3662--3670.

\end{thebibliography}

\begin{IEEEbiography}[{\fbox{\begin{minipage}[t][1.25in]{1in}%
Replace this box by an image with a width of 1\,in and a height of
1.25\,in!%
\end{minipage}}}]{Your Name}
 All about you and the what your interests are.
\end{IEEEbiography}

\begin{IEEEbiographynophoto}{Coauthor}
Same again for the co-author, but without photo
\end{IEEEbiographynophoto}

\end{document}